%\let\includefigures=\iftrue
%
% the following is to use blackboard bold fonts --
\let\useblackboard=\iftrue
%
% activate this if you don't have them.
%\let\useblackboard=\iffalse
%
% You might also need to remove this line.
\newfam\black

\input harvmac
%\input epsf.tex

%%BLACKBOARD FONT STUFF
\useblackboard
\message{If you do not have msbm (blackboard bold) fonts,}
\message{change the option at the top of the tex file.}
%Why is this magstep1? It makes the bb font bigger than rm --MBS
%You're asking me?  I stole this source code from your mom -- AEL.
%\font\blackboard=msbm10 scaled \magstep1
\font\blackboard=msbm10
\font\blackboards=msbm7
\font\blackboardss=msbm5
\textfont\black=\blackboard
\scriptfont\black=\blackboards
\scriptscriptfont\black=\blackboardss
\def\Bbb#1{{\fam\black\relax#1}}
\else
\def\Bbb{\bf}
\fi
%Macros for boxes from cordes moore ramgoolam
\def\boxit#1{\vbox{\hrule\hbox{\vrule\kern8pt
\vbox{\hbox{\kern8pt}\hbox{\vbox{#1}}\hbox{\kern8pt}}
\kern8pt\vrule}\hrule}}
\def\mathboxit#1{\vbox{\hrule\hbox{\vrule\kern8pt\vbox{\kern8pt
\hbox{$\displaystyle #1$}\kern8pt}\kern8pt\vrule}\hrule}}

%% for sub sub sections("...poor devil of a sub-sub...")
\def\subsubsec#1{\ifnum\lastpenalty>9000\else\bigbreak\fi
\noindent{\it{#1}}\par\nobreak\medskip\nobreak}
\def\yboxit#1#2{\vbotx{\hrule height #1 \hbox{\vrule width #1
\vbox{#2}\vrule width #1 }\hrule height #1 }}
\def\fillbox#1{\hbox to #1{\vbox to #1{\vfil}\hfil}}
\def\ybox{{\lower 1.3pt \yboxit{0.4pt}{\fillbox{8pt}}\hskip-0.2pt}}
%
%

%''We will all char together when we char
%And let there be no moaning at the bar''
%Tom Lehrer

\def\tPhi{{\tilde{\Phi}}}

\def\O{{\cal O}}

\

%  draw box of size #1pt and line thickness #2pt
\def\drawbox#1#2{\hrule height#2pt 
        \hbox{\vrule width#2pt height#1pt \kern#1pt \vrule width#2pt}
              \hrule height#2pt}
% Young tableaux

\def\Asym#1#2{\vcenter{\vbox{\drawbox{#1}{#2}
              \kern-#2pt       % line up boxes
              \drawbox{#1}{#2}}}}

%%More math defs
\def\d{\partial}
\def\p{\partial}

\def\frac#1#2{{#1 \over #2}}

%%bars

%hats
\def\hg{\hat{g}}
\def\hR{\hat{R}}
\def\hd{\hat{\d}}

%%misc

  % It's cf. not c.f. since `conferre' (latin) is one word.
 % These abbrevs. are so common that they're not italicized. 

\def\tPhi{{\tilde \Phi}}
\def\bZ{{\bar Z}}
\def\tS{{\tilde S}}

%% Refs
%Polyakovism
%\PolyakovRD
\lref\PolyakovRD{
A.~M.~Polyakov,
``Quantum Geometry Of Bosonic Strings,''
Phys.\ Lett.\ B {\bf 103}, 207 (1981).}
%%CITATION = PHLTA,B103,207;%%;
%\PolyakovRE
\lref\PolyakovRE{
A.~M.~Polyakov,
``Quantum Geometry Of Fermionic Strings,''
Phys.\ Lett.\ B {\bf 103}, 211 (1981).
%%CITATION = PHLTA,B103,211;%%
}
%\CooperVG
\lref\CooperVG{
 A.~R.~Cooper, L.~Susskind and L.~Thorlacius,
 ``Two-dimensional quantum cosmology,''
 Nucl.\ Phys.\ B {\bf 363}, 132 (1991).
 %%CITATION = NUPHA,B363,132;%%
}
%\SchmidhuberBV
\lref\SchmidhuberBV{
C.~Schmidhuber and A.~A.~Tseytlin,
``On string cosmology and the RG flow in 2-d field theory,''
Nucl.\ Phys.\ B {\bf 426}, 187 (1994)
[arXiv:hep-th/9404180].
%%CITATION = HEP-TH 9404180;%%
}
%EVA
\lref\evarg{Eva's supercritical RG paper}

%\FreedmanWX
\lref\FreedmanWX{
  D.~Z.~Freedman, M.~Headrick and A.~Lawrence,
  ``On closed string tachyon dynamics,''
  Phys.\ Rev.\  D {\bf 73}, 066015 (2006)
  [arXiv:hep-th/0510126].
  %%CITATION = PHRVA,D73,066015;%%
}
%\AdamsSV
\lref\AdamsSV{
  A.~Adams, J.~Polchinski and E.~Silverstein,
  ``Don't panic! Closed string tachyons in ALE space-times,''
  JHEP {\bf 0110}, 029 (2001)
  [arXiv:hep-th/0108075].
  %%CITATION = HEP-TH 0108075;%%
}
%\SilversteinQF
\lref\SilversteinQF{
  E.~Silverstein,
  ``Dimensional mutation and spacelike singularities,''
  Phys.\ Rev.\  D {\bf 73}, 086004 (2006)
  [arXiv:hep-th/0510044].
  %%CITATION = PHRVA,D73,086004;%%
}

%%Ian and friend
%\SwansonDT
\lref\SwansonDT{
  I.~Swanson,
  ``Cosmology of the closed string tachyon,''
  Phys.\ Rev.\  D {\bf 78}, 066020 (2008)
  [arXiv:0804.2262 [hep-th]].
  %%CITATION = PHRVA,D78,066020;%%
}
%\HellermanZZ
\lref\HellermanZZ{
  S.~Hellerman and I.~Swanson,
  ``A stable vacuum of the tachyonic E8 string,''
  arXiv:0710.1628 [hep-th].
  %%CITATION = ARXIV:0710.1628;%%
}
%\HellermanHF
\lref\HellermanHF{
  S.~Hellerman and I.~Swanson,
  ``Cosmological unification of string theories,''
  JHEP {\bf 0807}, 022 (2008)
  [arXiv:hep-th/0612116].
  %%CITATION = JHEPA,0807,022;%%
}
%\HellermanYM
\lref\HellermanYM{
  S.~Hellerman and I.~Swanson,
  ``Supercritical N = 2 string theory,''
  arXiv:0709.2166 [hep-th].
  %%CITATION = ARXIV:0709.2166;%%
}
%\HellermanFF
\lref\HellermanFF{
  S.~Hellerman and I.~Swanson,
  ``Dimension-changing exact solutions of string theory,''
  JHEP {\bf 0709}, 096 (2007)
  [arXiv:hep-th/0612051].
  %%CITATION = JHEPA,0709,096;%%
}
%\HellermanNX
\lref\HellermanNX{
  S.~Hellerman and I.~Swanson,
  ``Cosmological solutions of supercritical string theory,''
  Phys.\ Rev.\  D {\bf 77}, 126011 (2008)
  [arXiv:hep-th/0611317].
  %%CITATION = PHRVA,D77,126011;%%
}
%\HellermanQA
\lref\HellermanQA{
  S.~Hellerman and X.~Liu,
  ``Dynamical dimension change in supercritical string theory,''
  arXiv:hep-th/0409071.
  %%CITATION = HEP-TH 0409071;%%
}

%Others
%\MartinecTZ
\lref\MartinecTZ{
  E.~J.~Martinec,
  ``Defects, decay, and dissipated states,''
  arXiv:hep-th/0210231.
  %%CITATION = HEP-TH/0210231;%%
}
%\TseytlinPQ
\lref\TseytlinPQ{
  A.~A.~Tseytlin,
  ``String Vacuum Backgrounds With Covariantly Constant Null Killing Vector And
  2-D Quantum Gravity,''
  Nucl.\ Phys.\  B {\bf 390}, 153 (1993)
  [arXiv:hep-th/9209023].
  %%CITATION = NUPHA,B390,153;%%
}
%\TseytlinVA
\lref\TseytlinVA{
  A.~A.~Tseytlin,
  ``A Class of finite two-dimensional sigma models and string vacua,''
  Phys.\ Lett.\  B {\bf 288}, 279 (1992)
  [arXiv:hep-th/9205058].
  %%CITATION = PHLTA,B288,279;%%
}
%\CallanIA
\lref\CallanIA{
  C.~G.~Callan, E.~J.~Martinec, M.~J.~Perry and D.~Friedan,
  ``Strings In Background Fields,''
  Nucl.\ Phys.\  B {\bf 262}, 593 (1985).
  %%CITATION = NUPHA,B262,593;%%
}
%\PolchinskiRQ
\lref\PolchinskiRQ{
  J.~Polchinski,
  ``String theory. Vol. 1: An introduction to the bosonic string,''
%\href{http://www.slac.stanford.edu/spires/find/hep/www?irn=4634799}{SPIRES entry}
{\it  Cambridge, UK: Univ. Pr. (1998) 402 p}
}
%\ZamolodchikovTI
\lref\ZamolodchikovTI{
  A.~B.~Zamolodchikov,
  ``Renormalization Group and Perturbation Theory Near Fixed Points in
  Two-Dimensional Field Theory,''
  Sov.\ J.\ Nucl.\ Phys.\  {\bf 46}, 1090 (1987)
  [Yad.\ Fiz.\  {\bf 46}, 1819 (1987)].
  %%CITATION = YAFIA,46,1819;%%
}
%\AharonyRA
\lref\AharonyRA{
  O.~Aharony and E.~Silverstein,
  ``Supercritical stability, transitions and (pseudo) tachyons,''
  Phys.\ Rev.\  D {\bf 75}, 046003 (2007)
  [arXiv:hep-th/0612031].
  %%CITATION = PHRVA,D75,046003;%%
}

%\WittenGJ
\lref\WittenGJ{
  E.~Witten,
  ``Instability Of The Kaluza-Klein Vacuum,''
  Nucl.\ Phys.\  B {\bf 195}, 481 (1982).
  %%CITATION = NUPHA,B195,481;%%
}

%\HeadrickSA
\lref\HeadrickSA{
  M.~Headrick,
  ``A note on tachyon actions in string theory,''
  Phys.\ Rev.\  D {\bf 79}, 046009 (2009)
  [arXiv:0810.2809 [hep-th]].
  %%CITATION = PHRVA,D79,046009;%%
}

%\AlvarezGaumeWW
\lref\AlvarezGaumeWW{
  L.~Alvarez-Gaume and P.~H.~Ginsparg,
  ``Finiteness Of Ricci Flat Supersymmetric Nonlinear Sigma Models,''
  Commun.\ Math.\ Phys.\  {\bf 102}, 311 (1985).
  %%CITATION = CMPHA,102,311;%%
}

%\CecottiVB
\lref\CecottiVB{
  S.~Cecotti and C.~Vafa,
  ``Exact results for supersymmetric sigma models,''
  Phys.\ Rev.\ Lett.\  {\bf 68}, 903 (1992)
  [arXiv:hep-th/9111016].
  %%CITATION = PRLTA,68,903;%%
}
%\HorowitzBV
\lref\HorowitzBV{
  G.~T.~Horowitz and A.~R.~Steif,
  ``Space-Time Singularities in String Theory,''
  Phys.\ Rev.\ Lett.\  {\bf 64}, 260 (1990).
  %%CITATION = PRLTA,64,260;%%
}
%\HoravaYH
\lref\HoravaYH{
  P.~Horava and C.~A.~Keeler,
  ``Closed-String Tachyon Condensation and the Worldsheet Super-Higgs Effect,''
  Phys.\ Rev.\ Lett.\  {\bf 100}, 051601 (2008)
  [arXiv:0709.2162 [hep-th]].
  %%CITATION = PRLTA,100,051601;%%
}
%\HoravaHG
\lref\HoravaHG{
  P.~Horava and C.~A.~Keeler,
  ``M-Theory Through the Looking Glass: Tachyon Condensation in the $E_8$
  Heterotic String,''
  Phys.\ Rev.\  D {\bf 77}, 066013 (2008)
  [arXiv:0709.3296 [hep-th]].
  %%CITATION = PHRVA,D77,066013;%%
}
%\FreyKE
\lref\FreyKE{
  A.~R.~Frey,
  ``Backreaction in Closed String Tachyon Condensation,''
  JHEP {\bf 0808}, 053 (2008)
  [arXiv:0805.0570 [hep-th]].
  %%CITATION = JHEPA,0808,053;%%
}

\Title{\vbox{\baselineskip12pt
\hbox{BRX TH-606}\hbox{MIT-CTP 4051}}}
{\vbox{\centerline{Exact null tachyons from RG flows}}}
\vskip -2em
\centerline{Allan Adams${}^1\!$, Albion Lawrence${}^{2}\!$, and Ian Swanson${}^1\!$} 
\medskip
\centerline{${}^1$ {\it Center for Theoretical Physics, Massachusetts Institute of
Technology,}}
\centerline{{\it Cambridge, MA 02139}}
\centerline{${}^2$ {\it Theory Group, Martin Fisher School of
Physics, Brandeis University,}}
\centerline{{\it  MS057, 415 South St., Waltham, MA 02454}}

\bigskip
\noindent
We construct exact 2d CFTs, corresponding to closed string tachyon and metric profiles
invariant under shifts in a null coordinate, which can be constructed from any
2d renormalization group flow.  These solutions satisfy first order equations of motion
in the conjugate null coordinate.  The direction along
which the tachyon varies is identified precisely with the worldsheet scale,
and the tachyon equations of motion are the RG flow equations.

\medskip
%\draftmode
\Date{\number\day\ July 2009}

\newsec{Introduction}

It is an old idea that renormalization group flows in two-dimensional quantum field theories
can be lifted to time-dependent solutions of string theory. 
%or to nontrivial domain walls.   AA
When the RG flow describes the evolution of couplings
to relevant operators, the string theory background corresponds to a nontrivial tachyon
profile. This identification is precisely true in the limit of large dilaton slope
\refs{\CooperVG\SchmidhuberBV\SilversteinQF-\FreedmanWX}. Away from this limit,
the map between RG flows and real tachyon profiles is modified: in particular,
spacetime equations of motion are second order in time and space, 
while the RG flow equations are first order in scale 
\refs{\AdamsSV,\FreedmanWX}.\foot{That is, $[R,G] \neq 0$.} One may construct the
full spacetime profile in a derivative expansion for slowly varying tachyons or in conformal
perturbation theory for small tachyon expectation values \refs{\SchmidhuberBV,\FreedmanWX}.

In this work we pursue a modification of these arguments for closed string tachyons
in a background with a null shift symmetry.
% AA -- MAY CONFUSE THE READER, BETTER TO LEAVE IT FOR THE MAIN TEXT
% (apparently broken by a linear dilaton, though the
%action will transform under this shift only by a topological term, as we will see). 
% -- AA
It was pointed
out in \refs{\TseytlinPQ\TseytlinVA-\MartinecTZ} that for sigma models with such a symmetry, 
the equations of motion would be first order in the null directions and 
look more like renormalization group flows.  A wide class
of exact CFTs of this kind, with a timelike linear dilaton, have been worked 
out in \refs{\HellermanYM\HellermanZZ\HellermanHF\HellermanFF\HellermanNX\HoravaYH-
\HoravaHG}
(see also \refs{\HellermanQA\SwansonDT\FreyKE-\AharonyRA} for related studies).  
% AA --
%Here the beta functions receive no corrections beyond one-loop, due to the null shift symmetry
%and the relatively simple tachyon profile (much as the beta functions for the
%plane wave backgrounds \refs{\HorowitzBV}\ are one-loop exact).  
Due to the null shift symmetry
and the relatively simple tachyon profile,
the beta functions receive no corrections beyond one-loop in $\alpha'$
(much as the beta functions for the
plane wave backgrounds of \refs{\HorowitzBV}\ are one-loop exact).  
% -- AA
We generalize this work to 
describe null tachyon profiles with a null isometry and a timelike dilaton given any 
renormalization group flow.  In these flows, 
the direction along which the tachyon 
varies is mapped {\it precisely}\ to the worldsheet scale by a Lagrange multiplier constraint,
and the profile satisfies first order equations equivalent to the RG equations.

Note that these backgrounds will be exact in $\alpha'$, but not necessarily in $g_s$.  In particular,
we expect the dilaton to run to strong coupling in the past or future of these solutions.  
As usual, however, we can adjust the constant mode of the dilaton to push this strong coupling region as
far into the past or future as we wish.

\newsec{Worldsheet description of null tachyons}

In constructing our string theory background, we begin with
the tensor product of a two-dimensional target space and a conformal field
theory $\CC$. Assume this conformal theory has a set of local primary operators $\CO_a$.
Let us write the two-dimensional target space with the metric
\eqn\twodmet{
	ds^2 = - 2 dX^+ dX^-\ .
}
We now couple these theories by: 
\item{$\bullet$} Deforming $\CC$ by couplings $\int d^2\sigma \,u^a(X^+)\CO_a$ % AA -- ADDED SPACE -- AA
depending only on $X^+$.
\item{$\bullet$} Turning on a dilaton of the form $\Phi(X^+,X^-) = \gamma X^- + \tPhi(X^+)$.
The coefficient $\gamma$ is arbitrary, and can be shifted by rescaling $X^- \to \lambda X^-$,
$X^+ \to X^+/\lambda $; $\gamma$ has mass dimension one if $X^{\pm}$ has length dimension one. 
%AL have-->has

As in \refs{\HellermanZZ\HellermanHF\HellermanFF-\HellermanNX,\HellermanQA\SwansonDT-\FreyKE}, % AA-- WITH -> IN --AA
the dilaton  is linear in $X^-$.  The $X^+$ dependence of the tachyon will be nonlinear, as
the slope must shift between $X^+ = \pm \infty$ to make up for the change in the central charge 
associated with
the sector $\CC$ \refs{\HellermanQA,\FreedmanWX}. 
%Note that the string theory
%will not be weakly coupled everywhere in this background.  However, the constant mode will
%still be a modulus of this theory, so we can ensure that over any finite region of spacetime, 
%the dilaton has an upper bound as small as we like.

The full worldsheet action is\foot{In refs. \refs{\HellermanZZ\HellermanHF\HellermanFF-\HellermanNX,\HellermanQA\SwansonDT-\FreyKE}, there is also a term $(\d X^+)^2$ in the action corresponding
to a metric $G_{++}$ which is induced at large $X^+$.  This does not appear in our calculation.  We believe this amounts in part to a choice of scheme. A term $G_{++}(X^+)(\d X^+)^2$ can be removed
by a field redefinition $X^- \to X^- + f(X^+)$, where $\d_+ f = G_{++}$, without otherwise changing
the form of the action.  We would like to thank A. Frey for asking about this point.}
\eqn\fullact{
\eqalign{
	S_{\rm pert} = S_{\rm CFT} & + \frac{1}{4\pi\alpha'}
	\int d^2z \sqrt{g} \left[ - 2 g^{\alpha\beta} \d_{\alpha}X^+ \d_{\beta} X^-
		 \right. \cr
		& \ \ \ + \left.
		\alpha' \left(\gamma X^- + \tPhi(X^+)\right) R^{(2)} + u^a(X^+) \O_a \right]  \ .
}}
Note that $\CO_a$ will generally include the identity operator. Here $S_{\rm CFT}$ 
is the action for $\CC$.  However, our discussion will work just as
well if $\CC$ has no Lagrangian description.  In this latter case, the presence of
$S_{\rm CFT}$ merely denotes that for fixed $X^+,X^-$, the theory is the conformal
field theory $\CC$ perturbed by $\int d^2\sigma u^a(X^+) \O_a$.

The tachyon and metric couplings in \fullact\ have a symmetry under shifts of $X^-$.
This appears to be broken by the term in the dilaton linear in $X^-$.  However, note
that a shift in $X^-$ simply adds a total derivative (the Euler character of the worldsheet) 
to the action. Thus, this action 
%AA-- carries
respects
% --AA
an overall shift symmetry in $X^-$, which will be reflected
in the worldsheet beta functions.  
%AA--
Note, however, that the string genus expansion will not respect this shift symmetry, so one must treat the strong coupling region with caution. 
%--AA

We will assume that we know, in advance, complete information about
$S_{\rm CFT}$ perturbed by arbitrary couplings which might depend on the
worldsheet coordinates.  In particular, this means that we assume knowledge of
the full set of beta functions $\beta^q$ for $\CC$ perturbed by the terms
$\int d^2\sigma \bar{u}^a \CO_a$, where $\bar{u}$ are constant (c-number) couplings.
This includes the beta functions for the identity and for the dilaton. The dilaton beta function
is proportional to the Zamolodchikov c-function for the perturbed CFT, as we will discuss below.

The beta functions are also dependent on the contribution of the degrees of 
freedom of the perturbed theory $S_{\rm CFT} +  \frac{1}{4\pi\alpha'} \int d^2\!z\, u^a(X^+) \O_a$ %AA-- adjusted spacing --AA
to the beta function for the operator $(\p X^+)^2$, (e.g.,~for the spacetime metric 
$G_{++}$).  Even so, as we will argue below, this beta function is determined 
completely by the above information.

Finally, we will consider \fullact\ fixed to conformal gauge: namely, the worldsheet
metric is 
\eqn\cgmet{
	g_{\alpha\beta} = e^{2\phi}\hg_{\alpha\beta}\ .  
}
The partition function is given by
\eqn\partfn{
\eqalign{
	Z & = \int d\phi \bZ \ ,\cr
	\bZ & = \int DX^+ DX^- DY Db Dc\, e^{-S_{\rm pert}(X^+,X^-,Y) - S_{\rm FP}}\ ,
}}
Here $b,c$ are the conformal ghosts, and $S_{FP}$ is their action. $Y$ stands for the degrees of freedom on $\CC$.  
%AL -- Above was edited slightly (added the definition of $S_{FP}$, broke 1 sentence into 2.)
(Again, this is for ease of exposition: the central arguments of this work do not require that
$\CC$ have a Lagrangian description.) For a good string background, $\bZ$ must be
independent of $\phi$.

\newsec{Integrating out $X^-$}

In the action \fullact, $X^-$ appears as a Lagrange multiplier.   
By analytically continuing the theory to Lorentzian signature to do this integral,
we find that
\eqn\lmint{
	\bZ = \int DX^+ Db DcDY \delta(\sqrt{g} \p^2 X^+ - \frac{\gamma \alpha'}{2} \sqrt{g} R)
		e^{-\tS - S_{FP}} \ ,
}
%
%AL -- added S_{FP} to exponential
where 
\eqn\redact{
	\tS =  \int d^2\sigma \sqrt{g} \left[ \frac{1}{4\pi}
	\tPhi(X^+) R^{(2)} + u^a(X^+)\CO_a\right]\ .
}
In conformal gauge, we have
\eqn\cgcurv{
	\sqrt{g} R = \sqrt{\hg} \left(\hR - 2 \hd^2 \phi\right) \ ,
}
where $\hR$ is the 2d curvature for the fiducial metric $\hg$, and $\hd^2$ is the associated
Laplacian.  In two dimensions, $\sqrt{g} \d^2 = \sqrt{\hg}\hd^2$.
We can therefore write the delta function in \lmint\ as
\eqn\newdelt{
	\frac{1}{\det\left(\sqrt{g} \d^2\right)} \delta(X^+ + \gamma \alpha'\phi + Q) \ ,
}
where $Q$ depends only on the fiducial metric. Note that while $\sqrt{g}\d^2$ is Weyl invariant,
the determinant requires a regulator and thus 
contributes to the Weyl anomaly.  

The main lesson of this section is that for the action at hand, $X^+$ is identified 
precisely with worldsheet scale.
This is the physical basis of the observation below that for good string backgrounds,
% AA-- TYPO, U SHOULD BE LOWER CASE
%$U^a(X^+)$ 
$u^a(X^+)$ 
% --AA
will satisfy the first order RG equations, with $X^+$ functioning as the renormalization
group scale.

\newsec{Beta functions}

We wish to find the conditions under which $\frac{\delta}{\delta\phi(\sigma)} \bZ = 0$:
\eqn\varyz{
\eqalign{
	\frac{\delta}{\delta\phi(x)} \bZ  & = \int DX^+ DY Db Dc
	\left( \left[\delta_{\phi(x)} - (\delta_{\phi(x)}\tS)\right]\delta(X^+ + \gamma\alpha' \phi - Q)\right)
		\frac{1}{\det\p^2} e^{-\tS - S_{\rm FP}} \cr
	&\ \ \ \ \ \ \ \ + \delta^{(q)}_{\phi(x)} \bZ = 0 \ .
}}
Here $\delta^{(q)}_{\phi(x)}$ denotes the part of the variation induced by quantum effects:\foot{We
are using the conventions found in \refs{\FreedmanWX,\PolchinskiRQ}.}
\eqn\quantumvar{
	\delta^{(q)}_{\phi(x)} \bZ = \frac{1}{8\pi}  \beta^{q, \Phi}[u,\tPhi,\gamma] \sqrt{g}R^{(2)}
	+ \frac{1}{8\pi\alpha'} \beta^{q}_{++}[u,\tPhi,\gamma] \sqrt{g} (\p X^+)^2 + 
	\half \beta^{q,a}[u] \sqrt{g} \CO_a\ .
}
The quantum contribution $\beta^{q, \Phi}$ to the dilaton beta function comes 
from the determinant in \newdelt, and from the perturbations to $\CC$. It can thus be written as:
\eqn\leadingbf{
	\beta^{q,\Phi} = \frac{c[u] - 24}{6} \ ,
}
where $c[u]$ is the Zamolodchikov $c$-function for the theory $\CC$ perturbed
by the operators $u^a\CO_a$; the factor of 24 comes from the Fadeev-Popov ghosts
and from the determinant factor in \newdelt, \varyz\ that arose from integrating out $X^-$. 
The beta function $\beta^q_{++}$ comes from the 
contribution of this perturbed CFT to the beta function for the operator $(\p X^+)^2$.
Finally, $\beta^{q,a}$ are the contributions to the
beta functions for the couplings $u^a$, coming from this perturbed CFT.  
These are believed to be gradients of $c$, but to date there is no proof of that beyond
leading order in conformal perturbation theory \refs{\ZamolodchikovTI}.\foot{Note that this
proof breaks down at higher orders \refs{\FreedmanWX}. Furthermore, in at least
one case it is likely that the beta functions are not gradients of any function of the relevant
and marginal couplings alone \refs{\HeadrickSA}.}

We can exchange the $\phi$-derivative of the delta function in \varyz\ for an $X^+$
derivative and integrate by parts:
\eqn\deltader{
\eqalign{
	&  \int DX^+ DY Db Dc\, %AA-- added small spaces after measures --AA
	\delta_{\phi(x)}\delta(X^+ + \gamma\alpha' \phi - Q)\frac{1}{\det\p^2} e^{-\tS - S_{\rm FP}}\cr
	& \ \ \ \ \ = \gamma\alpha' \int DX^+ DY Db Dc \, \delta_{X^+}\delta(X^+ + \gamma\alpha' \phi - Q)
	\frac{1}{\det\p^2} e^{-\tS - S_{\rm FP}}\cr
	& \ \ \ \ \ = \int DX^+ DY Db Dc \, \delta(X^+ + \gamma\alpha' \phi - Q)\frac{1}{\det\p^2} e^{-\tS - S_{\rm FP}}\cr
	& \ \ \ \ \ \ \ \ \ \ \times
	\gamma\alpha'
	\left(\frac{1}{4\pi} \dot{\tPhi}(X^+(\sigma)) \sqrt{g} R + \dot{u}^a(X^+) \CO_a(\sigma) \right) \ ,
}}
where the dots denote derivatives with respect to $X^+$. Next, the classical variation of $\tS$
with respect to $\phi$ is:
\eqn\classdilvar{
\eqalign{
	\delta_{\phi(x)} \tS & = - \frac{1}{2\pi} \sqrt{g} \p^2\tPhi \cr
		& = - \frac{1}{2\pi} \sqrt{g} \left(\ddot{\tPhi}(X^+) (\p X^+)^2 + \dot{\tPhi}\p^2X^+
			\right) \ ,
}}
where we have used \cgcurv\ in the first line.
This expression will multiply the delta function in \varyz. 
As applied to eqn.~\varyz, the second term in \classdilvar\ can therefore be replaced by
\eqn\replacedbycurv{
	\frac{\gamma\alpha'}{4\pi} \dot{\tPhi} \sqrt{g} R\ ,
}
i.e.,~by a contribution to the dilaton beta function.

Collecting all of the terms in \varyz, we find that the variation of $\bZ$ vanishes if
\eqn\betafns{
\eqalign{
	& 2 \gamma \alpha' \dot{u}^a(X^+) +  \beta^{q,a} = 0\ , \cr
	& 4\gamma \alpha' \dot{\tPhi} + \beta^{q,\tPhi} = 0\ , \cr
	& - 4 \alpha' \ddot{\tPhi}(X^+) + \beta_{++} = 0\ .
}}
The left hand sides are the beta functions for the full theory \fullact. 
%AL -- removed "coupled to 2d gravity".
The first line
is the full beta function for $\CO$, the second line
line is the dilaton beta function, and the final line is the beta function for $(\p X^+)^2$. 

We claim that the $\beta^{q}$ arise entirely from the divergences in the perturbed CFT
due to singular operator products of the $\CO^a$, so that
as functions of $u$, $\beta^q$ are the same as for constant couplings. This
is true if there are no additional divergences arising from contractions of $X^+$.
(If $X^+$ was instead timelike, for example, such divergences would appear \refs{\FreedmanWX}.)
If the beta functions are computed perturbatively, this should be the result of the
diagrammatic arguments given in \refs{\HellermanYM\HellermanZZ\HellermanHF\HellermanFF-\HellermanNX}.  We have two additional arguments.

One is to note that there are no $X^+ X^+$ correlators by the following argument.  The delta function
in \lmint\ equates 
%AA-- added superscript
%$X$ 
$X^{+}$ 
%--AA
to the scale factor, $\phi$, plus a constant.  But when the beta functions
vanish, the integrand $\bar{Z}$ is completely independent of $\phi$, and so the two-point
function $\langle X^+ (\sigma)X^+(\sigma') \rangle$ is independent of $\sigma,~\sigma'$ (the factor $Q$ 
can be absorbed into $\phi$). While the beta functions are computed away from the conformal
point, this means that any divergences from contractions of $X^+$ must vanish at the conformal
point, and do not yield independent terms in the beta function equations.

Another argument is as follows.  It would appear that the path integral \lmint, together with
\redact, allows for nonvanishing OPEs of $X^+$, since the delta function makes the dilaton coupling
equivalent (after integrating by parts) to a standard kinetic term for $X^+$,
\eqn\candidatekin{
	S_{{\rm kin,} \Phi} \propto \int d^2\sigma \dot{\tilde{\Phi}} (\d X^+)^2 \ .
}
Near the RG fixed points at $X^+ = \pm\infty$, such a term is also induced
from integrating out the degrees of freedom in $\CC$ (see for example
\refs{\HellermanFF}), which cancels the dilaton contribution. As one flows
in $X^+$, the second line of \betafns\ indicates that the additional contributions to 
$G_{++}$ from $\CC$ are cancelled by the dilaton contribution.  Again,
one should be careful since the beta functions are computed away from the conformal
point.  As above, however, the additional divergences from $X^+ X^+$ correlators
should not give any additional contributions to the beta function equations describing the
conformal point itself.

We should also make sure that the arguments in this section and in \S3\ hold for
correlation functions -- that is, that the beta functions which we compute hold
for the Callan-Symanzik equation for correlation functions of local operators.\foot{We would
like to thank Joe Polchinski for reminding us of this issue.} The essential point is that
for local correlators (string scattering amplitudes are finite-dimensional
integrals of local correlators; recall also that three such operators are used to fix the 
$SL(2,\Bbb{C})$ invariance of the worldsheet CFT), the delta function constraint \newdelt\ should
only be modified at the point of the operator insertions. For example, if the worldsheet
was regulated with a lattice cutoff, one would integrate over all vaues of $X^-$ point by point.
Therefore, the delta function \newdelt\ is only modified at the operator insertion points.  
In the Callan-Symanzik equations, modifications
of the scale transformations which occur at the points of operator insertions give contributions
to the anomalous dimensions of operators, rather than to the 
beta functions.\foot{Similarly, if the beta functions are 
constrained due to a global symmetry, the constraints 
{\it on the beta functions}\ remain even when one computes correlation functions of local
operators that transform nontrivially under this global symmetry.}

Finally, eqs.~\betafns\ appear to have more equations than unknowns.  This is typical of
the beta function equations in string theory.  
In fact, the left hand side of the second equation in \betafns\ is conserved as a consequence of the
first and third equations: the vanishing of the beta functions on a flat 2d worldsheet
is sufficient to ensure that the theory is a CFT. One may then set it to zero by adjusting the
linear term in $\tPhi$.  This fact has been checked at one loop for sigma 
models in \refs{\CallanIA}, and for tachyons in \refs{\FreedmanWX}.  Furthermore, one
can compute $\beta_{++}^q$ to leading order in $X^+$ derivatives following the
discussion in \refs{\FreedmanWX}.  Combined with the leading-order result \refs{\ZamolodchikovTI}
\eqn\cfunction{
	\p_a c[u] = 24\pi^2 g_{ab} \beta^b[u]\ ,
}
one can show that the third line in \betafns\ follows from the first two lines, specifically
by taking the derivative of the second line.  More generally, one can argue that the vanishing
of the first two equations in \betafns\ implies the vanishing of the last equation, as follows. 
The Wess-Zumino condition implies that the derivative of the second term with respect to 
$X^+$ is proportional
to a linear combination of the beta functions of the theory.  Indeed, it should be a linear
combination of {\it all} of the beta functions of the theory; if any one relevant or marginally
relevant coupling is turned on, the theory will flow, and $\beta^{q,\Phi}$ will cease to be constant.
Thus, if one sets the beta function for all $G_{++}$ to zero, the derivative of the dilaton beta function
will be proportional to $G_{++}$.

It is worth noting that the beta function $\beta^q_{++}$ is determined by consistency of \betafns: taking
the derivative of the second equation, and using the first equation together with \leadingbf, we find that
\eqn\metricbf{
	\beta^q_{++} = \frac{1}{12\gamma^2\alpha'} \beta^{q,a}\p_a c\ .
}
It would be interesting to prove this directly. The $\gamma$ dependence arises
from the relationship between $X^+$ and the scale factor $\phi$, induced by integrating
out $X^-$.

%AA-- EDITED SENTENCE
%The final result of this section is that the spacetime for a null tachyon is completely 
%determined by {\it first}-order equations, and the dependence of the tachyon on $X^+$
%is identical to the RG flow in the scale $\phi = X^+/(\gamma\alpha')$.
The final result is that the spacetime evolution generated by a null tachyon profile 
is completely determined by the {\it first}-order equations of the associated RG flow, with the dependence of the 
tachyon on $X^+$ identical to that of the renormalized coupling on the RG scale $\phi = X^+/(\gamma\alpha')$.
%--AA

\newsec{Conclusions}

We have shown that any RG flow arising from 
%AA-- a-->the
%a
the 
%%--AA
perturbation of a CFT by a relevant
operator 
%AA-- edited for truthiness
%generates a null tachyon profile,
also defines an {\it exact} CFT describing the spacetime evolution of a null tachyon condensate,
%--AA
and that the resulting tachyon and dilaton profiles satisfy
{\it first}-order equations determining their evolution along 
%AA-- a-->the
%a
the 
%%--AA
null direction.  This generalizes
the work of \refs{\HellermanYM\HellermanZZ\HellermanHF\HellermanFF\HellermanNX\HoravaYH\HoravaHG
\HellermanQA\SwansonDT\FreyKE-\AharonyRA}.  As pointed out in \refs{\HellermanFF},
these are stringy ``bubbles of nothing'' (when the coupling to the identity operator in ${{\cal C}}$
flows)\foot{We would like to thank Simeon Hellerman for reminding us of this.}
--  analogs of \refs{\WittenGJ} -- in which dimensions
of spacetime are destroyed by an expanding domain of tachyon condensate
satisfying first-order equations of motion.  It would be interesting to study further examples,
such as the null tachyon generated from the 
RG flow of the $\Bbb{CP}^n$ model.\foot{We
would like to thank Eva Silverstein for suggesting this example.}
In this example, the K\"ahler class is known to flow precisely logarithmically with RG scale 
\refs{\AlvarezGaumeWW,\CecottiVB}, and it will therefore evolve linearly in $X^+$.  
This flow goes from a $c = 2n$ sigma model in the UV to a trivial $c = 0$ 
Landau-Ginzburg theory in the IR \refs{\CecottiVB}.

We should note that there often exists a scheme in which the tachyon
beta functions can be linearized, unless the tachyons mix nontrivially under the OPEs,
and are marginally relevant or satisfy some kind
of resonance condition (see \refs{\FreedmanWX}\ for a review and discussion).  
When the beta functions can be linearized, the $X^+$ dependence of the tachyon will be 
a simple exponential.  Furthermore, unlike the timelike examples in \refs{\FreedmanWX}, 
$X^+$ is identified precisely with the RG scale
so that the IR fixed point will be reached only at null infinity.  In this scheme there is less to be lost.  
When there are universal higher-order terms in the beta function,
the $X^+$ dependence will of course be more complicated.

\vskip .2cm
\centerline {\bf Acknowledgements}

We would like to thank A. Frey, M. Headrick, S. Hellerman, J. Polchinski, and E. Silverstein for 
discussions and comments on previous drafts.
%AA-- ADDED MY ACKS FOR KITP/ASPEN
A.A.~and A.L.~would like to thank the Kavli Institute for Theoretical Physics at UCSB 
%``Quantum Criticality and the AdS/CFT Correspondence'' Miniprogram 
and the Aspen Center for Physics 
%``String Duals of Finite Temperature and Low-Dimensional Systems''
for their hospitality while this work was completed.
The work of A.A.~and I.S.~is supported in part by funds provided by the U.S.~Department of 
Energy (D.O.E.) under cooperative research agreement DE-FG0205ER41360.
A.L.~is supported in part by DOE Grant  No.~DE-FG02-92ER40706, by a DOE 
Outstanding Junior Investigator award, and by the National Science Foundation 
under NSF grant NSF PHY05-51164.

\listrefs
\end